\documentclass[compsoc,conference,a4paper,10pt,times]{IEEEtran}
\IEEEoverridecommandlockouts
\usepackage{graphicx}
\usepackage[utf8]{inputenc}

\usepackage{float}
\usepackage{tikz,tkz-tab,pgf}
\usepackage{tikz}
\usepackage{pgfcore}
\usepackage{caption}
\usepackage{wrapfig}
\usepackage{multirow}
\usepackage{pgf-umlsd}
\usepackage{mathtools}
\usepackage{adjustbox}
\usepackage{graphicx}
\usepackage{enumitem}
\usepackage{booktabs}
\usepackage{epstopdf}
\usepackage{soul}

\usepackage[linesnumbered,ruled,vlined]{algorithm2e}
\usepackage{amsmath,amsthm,amssymb,amsfonts}
\definecolor{darkred}{rgb}{0.75,0,0}

\usepackage{makecell}
\usepackage{hyperref}

\usepackage[group-separator={,}]{siunitx}
\usetikzlibrary{positioning}
\usepackage{listings}
\usepackage{color}
\definecolor{lightgray}{rgb}{.9,.9,.9}
\definecolor{darkgray}{rgb}{.4,.4,.4}
\definecolor{purple}{rgb}{0.65, 0.12, 0.82}

\lstdefinelanguage{JavaScript}{
  keywords={constructor, mapping, uint256, bool,contract, public, typeof, new, true, false, catch, function, return, null, catch, switch, var, if, in, while, do, else, case, break},
  keywordstyle=\color{blue}\bfseries,
  ndkeywords={uint256,require, block.number,class, export, boolean, throw, implements, import, this},
  ndkeywordstyle=\color{darkgray}\bfseries,
  identifierstyle=\color{black},
  sensitive=false,
  comment=[l]{//},
  morecomment=[s]{/*}{*/},
  commentstyle=\color{purple}\ttfamily,
  stringstyle=\color{red}\ttfamily,
  morestring=[b]',
  morestring=[b]"
}

\theoremstyle{definition}
\newtheorem{definition}{Definition}[section]

\theoremstyle{remark}
\newtheorem{condition}{Condition}[section]

\def\tablename{Table}

\begin{document}

% The "title" command has an optional parameter, allowing the author to define a "short title" to be used in page headers.
%\title{Blockchain Currency Takeover}
%\title{On Replacing a Blockchain Currency with Metatransactions}
\title{On Blockchain Metatransactions}
%\author{István András Seres\inst{1}\orcidID{0000-0003-0143-4057} \and
%Arthur Gervais\inst{2}\orcidID{1111-2222-3333-4444}}
% The "author" command and its associated commands are used to define the authors and their affiliations.
% Of note is the shared affiliation of the first two authors, and the "authornote" and "authornotemark" commands
% used to denote shared contribution to the research.

%\author{Arthur Gervais}
%\affiliation{\institution{Imperial College London}}
%\email{a.gervais@imperial.ac.uk}

%\institute{Eötvös Loránd University
%\and
%Imperial College London}

%\institute{Princeton University, Princeton NJ 08544, USA \and
%Springer Heidelberg, Tiergartenstr. 17, 69121 Heidelberg, Germany
\author{\IEEEauthorblockN{István András Seres}
\IEEEauthorblockA{Department of Computer Algebra, Eötvös Loránd University\\
Email: istvanseres@caesar.elte.hu}}
\maketitle

\begin{abstract}
In cryptocurrencies, transaction fees are typically exclusively paid in the native platform currency. This restriction causes a wide range of challenges, such as deteriorated user experience, mandatory rent payments by decentralized applications, and blockchain community rivalries (e.g., coinism). Ideally, in a truly permissionless blockchain, transaction fees should be payable in any other cryptocurrency via so-called \emph{metatransactions}. In this paper, we formalize metatransactions, review existing ideas, and describe novel metatransaction design approaches. Under the assumption of sufficient market liquidity, we argue that metatransactions do not lower the security of cryptocurrency platforms. However, without changing the underlying blockchain, metatransaction designs typically increase transaction costs and reduce the blockchain transaction throughput.

% Keywords. The author(s) should pick words that accurately describe the work being
% presented. Separate the keywords with commas.
%\keywords{Blockchain, Cryptocurrency, Metatransaction, Security}
\end{abstract}

% The acknowledgments section is defined using the "acks" environment (and NOT an unnumbered section). This ensures
% the proper identification of the section in the article metadata, and the consistent spelling of the heading.
%\begin{acks}
%We are thankful to everyone!
%\end{acks}
%wjk - rewritten (hope you like it!); available on Skype (william.knottenbelt) if you want to discuss anything

\section{Introduction}
Cryptocurrencies have attracted significant attention, in part due to their ability to promote neutrality through disintermediation.
Their neutrality stems from their typically \emph{permissionless} nature: anyone is allowed to join and leave the network at any time, while their \emph{censorship resilience} protects individuals from potentially overreaching entities. Another aspect of neutrality is the inherent ability to fork an open-source blockchain, i.e.\ to clone one blockchain into two independent versions, if e.g.\ the blockchain community's vision diverges. This freedom, however, comes at a price. Instead of unifying efforts, we have witnessed an explosion in the number of blockchains and cryptocurrency tokens that compete rather than collaborate. This competition naturally arises as many blockchains have the same objective; to become the predominant decentralised platform for asset transfers and decentralised application hosting.

While competition in open-source projects and their ability to fork are not new phenomena (see e.g.\ Linux), in the context of blockchains, financial incentives exacerbate competition. We count historically about 900 Linux distributions\footnote{Source: \url{https://distrowatch.com/}} within the last 28 years, and the last 10 years have brought to fruition circa \num[group-separator={,}]{900} blockchain coins and about \num{1400} blockchain tokens\footnote{Source: \url{https://coinmarketcap.com/coins/views/all/}} deployed on existing blockchains.

All blockchains we are aware of rely on one native currency $\mathcal{C}$. $\mathcal{C}$ e.g. serves as the currency to pay transaction fees to miners and may be minted when creating a block. Typically, no other currency $\mathcal{C}^{*}$ is eligible to replace $\mathcal{C}$ in this special function. Indeed some blockchain creators openly criticise the use of alternative currencies\footnote{\url{https://github.com/ethereum/EIPs/blob/master/EIPS/eip-1559.md}}. We observe that this exclusivity leads to several challenges:

\begin{description}
\item[Deteriorated User Experience] A user that wishes to perform a transaction that does not involve $\mathcal{C}$, e.g.\ a stablecoin transaction\footnote{Stablecoins are typically less volatile coins that are, for example, pegged to a fiat currency.}, such as Tether or DAI~\cite{berentsen2019stablecoins}, is required to purchase and use $\mathcal{C}$ to perform the transaction. This results in additional usability frictions and UI complexities, which could be avoided if transaction fees could be paid directly via another currency $\mathcal{C}^{*}$. 
% wjk comment
%\hl{When we talk about miners being paid are we talking about block rewards or transaction fees? It sounds like the latter more than the former?}
\item[Mandatory Rent] Every project $\mathcal{P}$ that builds upon an existing blockchain is forced to adopt $\mathcal{C}$ and implicitly pays rent to the holder of $\mathcal{C}$. That is because users of $\mathcal{P}$ are required to purchase $\mathcal{C}$ whenever they want to interact with $\mathcal{P}$.
\item[Social Coinism] A few vocal owners of $\mathcal{C}$ publicly become the advocates of this coin, despite the availability of other potentially more suitable technologies and blockchains. Coinism is likely to alienate communities, preventing healthy, productive collaborations and could ultimately stifle innovation.
%\item[Potentially Unfair Economical Wealth Distribution] Many blockchain coins are distributed via private and public coin offerings. Private agreements often favor the initial buyer, while the project creator retain a majority share of the circulating supply. Those that buy or mine $\mathcal{C}$ early, therefore amass disproportionate amounts of $\mathcal{C}$, potentially leading to a danger of market cornering~\cite{kondor2014rich}.
%\item[Privacy] Users compete in a transparent and public transaction fee auction market to confirm their transactions in a block. The lack of transaction fee privacy leads to consensus instability~\cite{carlsten2016instability,daian2019flash}.
\end{description}

To remedy these problems, in this work, \emph{we study the options available to replace the currency $~\mathcal{C}$ of an already deployed blockchain with one or several alternative currencies, and consider the implications}. We adopt the blockchain community jargon \emph{metatranscations} as denoting transactions that pay fees to miners or other intermediaries in a currency other than $\mathcal{C}$. Summarizing, we provide the following contributions within this work:
\begin{itemize}
    \item We review an existing fee delegation scheme that supports the payment of transaction fees in currencies other than a native blockchain currency $\mathcal{C}$.
    \item We formalize the notion of metatransactions and present two novel designs, one miner-based and one off the chain scheme.
    \item We discuss the implications of metatransactions and show how an adversary may perform a hostile blockchain takeover with metatransactions. We argue that metatransactions do not affect the blockchain's security under assumptions of market liquidity.
\end{itemize}

The rest of the paper is organized as follows. Section~\ref{sec:background} presents pertinent background on smart contract enabled blockchains. In Section~\ref{sec:formaldef} we formally define metatransactions and present an existing fee delegation design, while Section~\ref{sec:newschemes} shows two new metatransaction schemes. Section~\ref{sec:comparingproposals} compares these two metatransaction protocols. Section~\ref{sec:securityofmetatx} analyses the economic and security implications of introducing metatransactions into a blockchain system. We highlight related work in Section~\ref{sec:related-work} and point out possible extensions as future directions in Section~\ref{sec:extensions}.

\section{Background} \label{sec:background}
Blockchains enable the construction of append-only immutable ledger maintained by a distributed network of nodes~\cite{nakamoto2008bitcoin}. The majority of current systems~\cite{nakamoto2008bitcoin, wood2014ethereum} rely on a random leader election process as part of their consensus mechanism. An elected participant decides on the current state transition: a set of \emph{transactions} altering the ledger state, arranged in a so called \emph{block}.
For example, in Proof-of-Work (PoW) blockchains, the leader is the first participant to solve a computationally expensive puzzle~\cite{nakamoto2008bitcoin}. For a thorough background on Bitcoin and blockchain, we refer the reader to a systematization of knowledge~\cite{bonneau2015sok}. Smart contract enabled blockchains allow to encode programmable logic which may execute transactions and manipulate cryptocurrency amounts. Contract code is typically executed in a Virtual Machine (VM), which is a quasi-Turing complete execution environment executing bytecode. The number of computational steps a transaction can perform in the VM is bounded to deter denial-of-service (DoS) attacks~\cite{perez2019broken}.

A crucial aspect of a smart contract VM is the fee mechanism. Every VM opcode costs a certain fee which should ideally be proportional to the computational complexity of that opcode. There typically exist a ``base'' transaction fee representing the minimum transaction fee\footnote{The base fee in Ethereum is e.g., \num[group-separator={,}]{21000} gas (\num[group-separator={,}]{10000} in Tezos~\cite{goodman2014tezos}), and not applicable to internal transactions.}. Note that due to the halting problem~\cite{turing1937computable} one cannot anticipate the required fees of smart contract execution and therefore typically define a maximum fee. Once this maximum is reached, the execution is terminated. To the best of our knowledge, we are not aware of a blockchain that natively supports the payment of transaction fees in a different currency than the native blockchain currency. %In Ethereum, transaction fees are denominated in the gas that can only be purchased with the native currency. The maximum fees for a transaction are therefore determined with $\mathit{txFee} = \mathit{gasPrice} * \mathit{gasLimit}$.
\par\smallskip
\noindent\textbf{Off-Chain Transactions}
For use cases such as micropayments, on-chain transaction fees and confirmation time latencies represent significant practical obstacles. Payment channels allow parties to exchange transactions locally, i.e.\ not broadcasting them to the network, and rather update a local balance sheet. The blockchain is only utilized as a recourse for channel initiations, disputes and closures. This permits payment channel participants to send payments without on-chain fees and short confirmation times. We refer the reader to a systematization of knowledge~\cite{gudgeon2019sok} for an overview of off-chain constructions.

\subsection{Notation}
In the following, we denote $H$ as a cryptographically secure hash function. Let $sk$ and $pk$ denote secret and private keys respectively of an asymmetric cryptographic protocol. The corresponding blockchain address is obtained by $addr(pk)$. We refer to a specific tuple element by its name in the subscript. Transactions, defined in Section~\ref{sec:txdef}, can be broadcast on-chain or sent off-chain. To capture this, we introduce the functions $\mathtt{broadcast}(\cdot)$ and $\mathtt{sendOffChain}(\cdot)$. If necessary we indicate in the superscript of these functions the blockchain platform, where the said transaction is issued.

\section{Metatransactions}\label{sec:formaldef}
In this section, we provide a system model and formalize metatransactions. We also review existing fee delegation schemes. Informally, a metatransaction is a (set of) transaction(s) that allows paying transaction fees in a currency other than the native blockchain currency~$\mathcal{C}$.

\subsection{System Model}
In the following, we describe our system model, functionalities and threat model for metatransactions.

\subsubsection{Actors} The following actors may be involved in a metatransaction:
\begin{itemize}[leftmargin=*]
    \item \emph{Sender}: a blockchain user with an asymmetric key pair which issues metatransactions.
    \item \emph{Receiver}: a receiver of a metatransaction.
    \item \emph{Relayer}: a metatransaction design might involve one or more non-trusted intermediaries that assist in the transaction processing.
    \item \emph{Smart contract}: a contract to enforce the metatransaction protocol logic, e.g., to reimburse miners and/or intermediaries for their services. 
    \item \emph{Miner}: choose to accept (meta)transactions within a mined block.
%    \item \emph{P2P full nodes}: verify and forward transactions on the P2P network layer. They may decide to not relay metatransactions.
\end{itemize}
\subsubsection{Informal Functionality} A metatransaction protocol enables a sender to issue and pay for blockchain transactions without using  $\mathcal{C}$, the native currency of the blockchain. %Once a sender issues metatransactions in the blockchain P2P network, full nodes may verify and forward these until miners eventually include them in the blockchain.

\subsubsection{Communication model} We assume broadcast transactions in the P2P network are delivered with a maximum delay under the bounded synchronous communication setting~\cite{attiya2004distributed}. Furthermore, we assume all actors can access and read the current head of the blockchain to verify if transactions are appended to the blockchain. We remark that these are standard assumptions in the blockchain literature~\cite{badertscher2017bitcoin,garay2015bitcoin}.

\subsubsection{Threat model} We assume that the cryptographic primitives of the blockchain hosting the smart contract are secure. Furthermore, we assume the adversary cannot corrupt more than $x\%$ of consensus participants of the blockchain (i.e.\ $50\%$ of the computational power in case of a PoW blockchain). Note that under the presence of selfish mining attacks, this threshold may be lowered~\cite{eyal2018majority, gervais2016security}. The adversary may corrupt or perform DoS attacks on intermediaries. Lastly, we assume that senders are not eclipsed, i.e.\ they are capable of broadcasting transactions~\cite{marcus2018low,wust2016ethereum, gervais2015tampering}.

\subsubsection{Security goals}\label{sec:securitygoals}
Designing a practical and trust-minimising metatransaction protocol is non-trivial. For example, one cannot know a priori which miner is going to include a transaction within a blockchain block. This might require to allow \emph{anyone that mines a metatransaction} to claim the corresponding transaction fee in a non-native currency $\mathcal{C}^{*}$. Moreover, for security purposes, transactions composing a metatransaction must be executed atomically, i.e.\  either all transactions execute, or none. Atomicity ensures that the sender has to pay fees, and miners must perform their service if they collect fees. Finally, compared to ordinary transactions, metatransactions should not incur additional delays when being included in a blockchain and in particular should be resilient to censorship. These observations lead us to the following informal security goals that metatransactions should ideally satisfy:

\begin{description}
\item[Atomicity:] A metatransaction may be composed of multiple transactions (cf.\ Definition~\ref{def:tx}). To execute a metatransaction successfully, any sub-transaction of a metatransaction must be executed atomically, i.e., within the same block.
\item[Censorship Resistance:] Metatransactions should be as difficult to censor as regular transactions.
%\item[Economic Viability:] {\color{blue}@Kaihua}The cost of issuing a metatransaction should be of the same order as a regular transaction.
\end{description}

\subsection{Formal Definition}\label{sec:txdef}
We provide the following formalism building on related work~\cite{badertscher2017bitcoin,garay2015bitcoin} to capture metatransactions.
\theoremstyle{definition}
\begin{definition}{\textbf{Transaction}}\label{def:tx}
A transaction is a tuple $\mathbf{tx}=(\mathbf{s}, \mathbf{r}, \mathbf{txFee}_{\mathcal{C}}, \mathbf{txFee}_{\mathcal{C}^{*}}, \delta),$ where $\mathbf{s}$ is the sender and $\mathbf{r}$ is the receiver of the transaction. The transaction fee offered in the native currency is denoted as $\mathbf{txFee}_{\mathcal{C}}$ and $\mathbf{txFee}_{\mathcal{C}^{*}}$\ denotes the transaction fee offered in another currency $\mathcal{C}^{*}$ than the native currency $\mathcal{C}^{}$. $\delta$ represents a reference to another transaction. In particular, $\delta$ is the $\mathbf{id}$ of a transaction $\mathbf{tx}'$ which should be included in the blockchain before $\mathbf{tx}$ can be appended. We define the id of a transaction to be $\mathbf{id_{tx}}=H(\mathbf{tx})$. Let $\mathit{mined}(\mathbf{tx})$ be a predicate which is true if $\mathbf{tx}$ is appended to the blockchain and false otherwise. For the time being, we assume that all transactions are issued on the same blockchain.
\end{definition}

Our Definition~\ref{def:tx} captures the fact that in most cryptocurrencies, paying transaction fees in $\mathcal{C}^{*}$ can only be achieved implicitly through the side-effects of an additional transaction on the same blockchain (we discuss cross-chain metatransactions in Section~\ref{sec:xchainscheme}). Intuitively one can think of a metatransaction as one transaction which pays no transaction fees in $\mathcal{C}$, coupled with another transaction which pays the equivalent fees in $\mathcal{C}^{*}$. $\delta$ allows us to capture dependency relations between transactions issued on the same or different blockchains to construct atomic transactions. Atomicity is required to e.g., prevent an adversarial miner from stealing sender funds and similarly to prevent the sender from issuing transactions without paying due fees. We remark that Definition~\ref{def:tx} omits several subtleties of real-world cryptocurrency transactions: for instance, we assume but do not indicate that transactions are signed by the sender. We are not aware of any blockchain design that explicitly allows paying transaction fees in a currency different than the native currency.

%Therefore, we define metatransactions as transaction pairs: the first transaction performs a state transition (such as currency transfer, smart contract execution), while the second transaction rewards the miners for mining the transaction \hl{(on which blockchain does the first transaction execute, and which the second?)}. Importantly, both transactions must be executed atomically to prevent, e.g.\ an adversarial miner from stealing user funds or users from free-riding.

\begin{definition}{\textbf{Metatransaction}} \label{def:metatx}
A metatransaction $\mathbf{tx}_{meta}$ is a pair of transactions $(\mathbf{tx}_0,\mathbf{tx}_1)$, where $\mathbf{tx}_{0,\mathbf{txFee}_{\mathcal{C}}}=\mathbf{tx}_{0,\mathbf{txFee}_{\mathcal{C}^{*}}}=\mathbf{tx}_{1,\mathbf{txFee}_{\mathcal{C}}}=0$, $\mathbf{tx}_{1,\mathbf{txFee}_{\mathcal{C}^{*}}}\neq 0$ and $\mathbf{tx}_{1,\delta}=\mathbf{id_{tx_0}}$. We allow $\mathbf{tx}_0=\emptyset$, if a single transaction is sufficient to perform a metatransaction.
\end{definition} 

%is that most cryptocurrencies do not allow to perform actions and transfer non-native currencies in a single transaction. However, if it is allowed, the definition captures this case by letting $\mathbf{tx}_0=\emptyset$. 

On the avenue towards the first metatransaction designs, we review an existing fee delegation scheme.

\subsection{Relayer-based Fee Delegation Scheme}\label{sec:relayerschemes}
A fee delegation scheme allows a sender to delegate the payment of transaction fees in $\mathcal{C}$ to a non-trusted party, while receiver reimburses the non-trusted party in $\mathcal{C}^{*}$. In this scheme, ultimately, miners receive $\mathcal{C}$ for mining the sender's transaction~\cite{buterin2019Layer2Gas}. While such a scheme may improve usability, it does not solve the other aforementioned obstacles that motivate our work (e.g., mandatory rent, coinism).

\begin{figure}%[6]{l}{0pt}
\centering
\includegraphics[width=\linewidth,trim={2.3cm 12.4cm 8cm 4cm},clip]{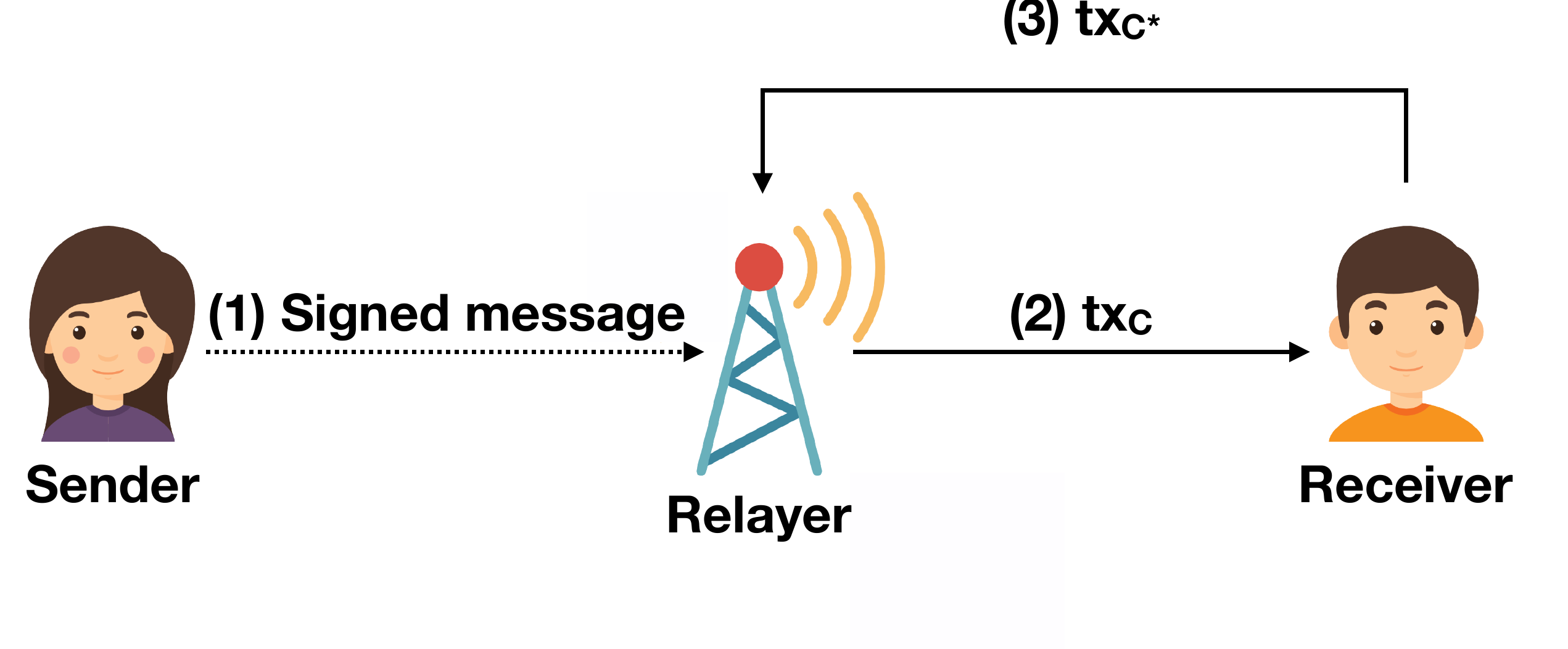}
\caption{A relayer fee delegation scheme (not a metatransaction design). The dotted arrow (1) represents a signed off-chain message, while continuous lines represent regular on-chain transactions. The relayer issues a regular on-chain transaction on behalf of the sender, and the relayer is compensated in a non-native currency $\mathcal{C}^{*}$. Note that in this example the transaction fees in $\mathcal{C}^{*}$ is borne by the receiver, i.e.\ the sender delegates the transaction fee payment to the receiver.}
\label{fig:relayerscheme}
\end{figure}

A fee delegation scheme involves a non-trusted intermediary, referred to as \emph{relayer}, which performs the conversion between $\mathcal{C}^{*}$ and $\mathcal{C}$ on behalf of the sender. On the blockchain, transactions, therefore, appear to be paid in $\mathcal{C}$. A possible scheme operates as follows (cf.\ Figure~\ref{fig:relayerscheme}). \emph{(1)} A sender transmits a signed message to the relayer off the blockchain. The signed message contains the to-be-invoked smart contract (or target address), function, corresponding arguments, etc. \emph{(2)} The relayer then creates an on-chain transaction $\mathbf{tx}_{\mathcal{C}}$ which contains this signed message. $\mathbf{tx}_{\mathcal{C}}$ pays transaction fees in $\mathcal{C}$. The execution of $\mathbf{tx}_{\mathcal{C}}$ triggers the execution of the second transaction $\mathbf{tx}_{\mathcal{C}^{*}}$, which reimburses the relayer for its services in $\mathcal{C}^{*}$. This reimbursement is paid by the receiver, cf. Figure~\ref{fig:relayerscheme}. The relayer, therefore, acts as an implicit exchange between $\mathcal{C}^{*}$ and $\mathcal{C}$. The sender may additionally incentivise the relayer for its services by paying a small fee. Note that such fee delegation schemes do not fall under metatransaction designs (cf.\ Definition~\ref{def:metatx}), because currency $\mathcal{C}$ remains required.

%The first class of metatransaction designs require an additional intermediary, the \emph{relayer}~\cite{buterin2019Layer2Gas}. A relayer is a non-trusted party who is financially motivated to bundle signed messages into regular transactions and to later broadcast them.
%An example design operates as follows: {\color{blue}sender} \hl{(A bit vague; who is the ``one'' exactly?)} {\color{blue}issues} a signed message to a relayer service out of band. Later these signed blobs of data are picked up by relayer nodes and are bundled into transactions, subsequently issued on-chain. Note that both on-chain transactions (2) and (3) are issued atomically, see Figure~\ref{fig:relayerscheme} \hl{(Nice pic! Would be good to know who is on C and who on C star though)}.

Positively, the execution of $\mathbf{tx}_C$ and $\mathbf{tx}_{C^{*}}$ is atomically enforced due to their explicit dependence. A major disadvantage of such fee delegation scheme, however, is that the relayer can censor a sender or become unavailable (e.g., due to a crash or a DoS attack). One avenue to mitigate such weaknesses could be the establishment of a network of relayers, we, however, leave this to future work. %To repel censorship attacks a sender might need to issue a regular transaction in the worst case, which would defeat the purpose of a relayer node. Furthermore, a relayer node can easily deanonymize transactions and link them to real-world users or one of their unique identifiers (IP address, e-mail etc.) 

\section{Metatransaction Designs} \label{sec:newschemes}

In the following section we outline our metatransaction design proposals.

\subsection{Miner-based Metatransaction}\label{sec:minerscheme}

Our first metatransaction design enables a sender $\mathcal{S}$ to pay transaction fees directly in $\mathcal{C}^{*}$ to a miner $\mathcal{M}$, without resorting to an intermediary (cf.\ Algorithm~\ref{alg:minermetatransaction}). The protocol proceeds as follows:
\begin{enumerate}
    \item $\mathcal{S}$ creates a transaction, $\mathbf{tx}_0 = (s_0, r_0, 0, 0, \emptyset)$ which executes a desired action (e.g.\ a cryptocurrency transfer or a smart contract invocation), and pays no transaction fees.
    \item A second transaction $\mathbf{tx}_1 = (s_0, \mathcal{A}(\mathcal{M}), 0, \mathit{txFee}_{\mathcal{C}^{*}}, \mathbf{id_{tx_0}})$ triggered by $\mathcal{S}$ or $\mathbf{tx}_0$ pays transaction fees for both $\mathbf{tx}_0$ and $\mathbf{tx}_1$ in $\mathcal{C}^{*}$ to a miner-can-spend address $\mathcal{A}(\mathcal{M})$.
    %\item $\mathcal{M}$ can claim accrued transaction fees in $\mathcal{C}^{*}$ at the pre-determined address by redeeming it with a publicly known secret key.
    \item $\mathcal{M}$ can claim accrued transaction fees in $\mathcal{C}^{*}$ from the miner-can-spend address.
\end{enumerate}

\begin{algorithm}
\SetAlgoLined
    $\mathbf{tx}_{meta}=(\mathbf{tx_0},\mathbf{tx_1})=((s_0,r_0,0,0,\emptyset),(s_0,addr(G),0,\mathit{txFee}_{\mathcal{C}^{*}},\mathbf{id_{tx_0}})$\;
   $\mathtt{broadcast}(\mathbf{tx}_{meta})$\;
   $\mathtt{broadcast}((addr(G),r_{miner},0,\mathit{txFee}_{\mathcal{C}^{*}},\mathbf{id_{tx_1}}))$\;
 \caption{A miner-based metatransaction construction. A sender creates two transactions, one which executes an action and doesn't pay fees, while the other pays fees in $\mathcal{C}^{*}$.}
 \label{alg:minermetatransaction}
\end{algorithm}

%, i.e.\ a miner cannot just include transaction $\mathbf{tx}_1$ to claim the transactions fees in since {\color{blue}transaction $\mathbf{tx}_1$ \hl{(what exactly?)} would contain an incorrect nonce of the corresponding sender account. We note this scheme can also be used in a non-backward compatible way, whereby transaction $\mathbf{tx}_1$ would simply be an internal transaction.  Moreover all three transactions must be included in the same block by the miner; otherwise, transaction (3) might be claimed in a subsequent block.

\begin{figure}%[14]{r}{0.65\textwidth}
\centering
\includegraphics[width=\linewidth,trim={5cm 6.4cm 6cm 8cm},clip]{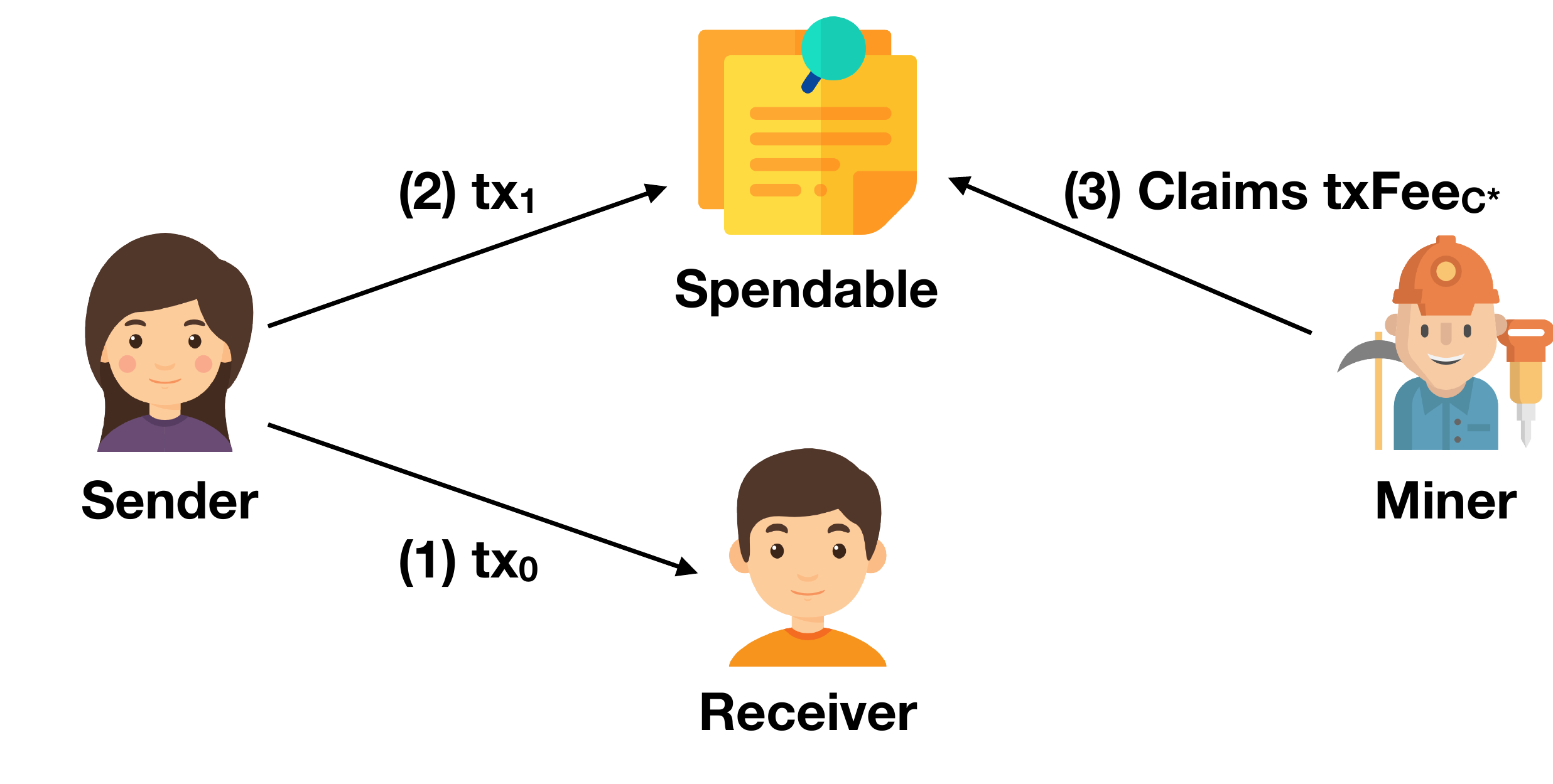}
\caption{The order of transactions involved in the miner-based metatransaction scheme. Sender transfers transaction fees in $\mathcal{C}^{*}$ to an anyone-can-spend address. Miners subsequently collect accrued fees at that address. Note, in this construction all transactions occur on-chain.}\label{fig:minerscheme}
\end{figure}

The challenge here is that a sender does not know beforehand which miner will mine the next block. A solution is to send the transaction fee to an anyone-can-spend address, which is likely to be claimed by the next miner. We could envision that the community agrees on an address for the sake of clarity and uniformity\footnote{For instance, the address corresponding to the base point $G$ of the secp256k1 curve. The corresponding secret key is publicly known, $sk=1$, meaning that any miner can easily claim accrued transaction fees in any given block.}. We remark that in some cryptocurrencies, one might be able to send fees directly to miners without knowing a priori who will mine said transaction (the Ethereum Virtual Machine provides a corresponding endpoint). This minimizes the added complexity of metatransactions, eliminating the need for an anyone-can-spend address.

Note that transaction $\mathbf{tx}_0$ and $\mathbf{tx}_1$ must be atomic, otherwise sender and miner could steal coins respectively. The atomicity of these transactions can be guaranteed, for instance, with a transaction counter.

Interestingly, to reduce on-chain costs, multiple $\mathbf{tx}_0$ can be batched, while one $\mathbf{tx}_1$ rewards the miner for their atomic inclusion. The upper bound for batching corresponds to the number of transactions that fit within a block.

\subsection{Payment-channel-based Metatransaction} \label{sec:channelscheme}
One bottleneck of the miner-based scheme introduced in Section~\ref{sec:minerscheme} is the necessity to reimburse miners with $\mathbf{tx}_1$ following each on-chain transaction $\mathbf{tx}_0$. This overhead inflates the blockchain and increases fees. To amortize on-chain costs, a sender might want to pay transaction fees off-chain (e.g.\ via payment channels) directly to miners.

In the following, we show how payment channels can facilitate metatransactions.
One might think of a payment channel as a balance sheet between the sender $\mathcal{S}$ and the miner $\mathcal{M}$.
By updating the balances of the two sides, $\mathcal{S}$ is able to reimburse $\mathcal{M}$ without touching the blockchain.
Interestingly, transaction fees are unidirectional (i.e.\ from $\mathcal{S}$ to $\mathcal{M}$), hence, we make use of simple Spilman-style payment channels~\cite{spilman2013anti}. Considering a channel $C_{\mathcal{S}\rightarrow\mathcal{M}}$, our proposed payment channel metatransaction scheme operates as follows:
\begin{description}
    \item[Channel Establishment]
    To establish $C_{\mathcal{S}\rightarrow\mathcal{M}}$, $\mathcal{S}$ necessarily needs to deposit upfront a sufficient amount of transaction fee collateral in $\mathcal{C}^{*}$.
    The initial channel balance can be described as a function defined as $[\mathcal{S}\rightarrow \mathsf{B}_{\mathcal{S}}^0,\mathcal{M}\rightarrow \mathsf{B}_{\mathcal{M}}^0]$, where $\mathsf{B}_{\mathcal{S}}^0 = \mathsf{B}_{C_{\mathcal{S}\rightarrow\mathcal{M}}}$ which equals the total funds locked in the channel and $\mathsf{B}_{\mathcal{M}}^0 = 0$.
    \item[Metatransaction Issuing]
    For every on-chain transaction $\mathbf{tx}_0^i$, $\mathcal{S}$ pays the transaction fee $\underline{\mathbf{tx}}^i_{1,\mathbf{txFee}_{\mathcal{C}^{*}}}$ to $\mathcal{M}$ through an off-chain transaction $\underline{\mathbf{tx}}_1^i$.\footnote{We denote an transaction that is off-chain by underlining it (i.e.\ $\underline{\mathbf{tx}}$).}
    To issue $\underline{\mathbf{tx}}_1^i$, $\mathcal{S}$ signs an aggregation transaction $\mathbf{tx}_{\sum_i}$ which accumulates all the valid off-chain payments. In other words, $\mathbf{tx}_{\sum_i}$ reflects the latest balance of $C_{\mathcal{S}\rightarrow\mathcal{M}}$, i.e.\ $[\mathcal{S}\rightarrow \mathsf{B}_{\mathcal{S}}^{i-1} - \underline{\mathbf{tx}}^i_{1,\mathbf{txFee}_{\mathcal{C}^{*}}} ,\mathcal{M}\rightarrow \mathsf{B}_{\mathcal{M}}^{i-1} + \underline{\mathbf{tx}}^i_{1,\mathbf{txFee}_{\mathcal{C}^{*}}}]$.
    %Note, $\underline{\mathbf{tx}}_1^i$ is valid \emph{iff} $\mathbf{tx}^i_0$ has been mined.
    In such manner, a metatransaction $\mathbf{tx}_{meta} = (\mathbf{tx}_0^i, \underline{\mathbf{tx}}_1^i)$ only requires a single on-chain transaction.
    \item[Channel Closure]
    Whenever the balance of $\mathcal{S}$ is depleted or $\mathcal{M}$ decides to finalize all the off-chain payments on-chain,
    $\mathcal{M}$ can close $C_{\mathcal{S}\rightarrow\mathcal{M}}$ by publishing the last aggregation transaction $\mathbf{tx}_{\sum_N}$. To avoid the funds locked forever because of an unresponsive miner, a lifetime is set for $C_{\mathcal{S}\rightarrow\mathcal{M}}$.
    If $\mathcal{M}$ doesn't close $C_{\mathcal{S}\rightarrow\mathcal{M}}$ before it expires, $\mathcal{S}$ can then close the channel and takes back all the funds.
    %We emphasize that, to guarantee atomicity, $\mathbf{tx}_{\sum_N}$ \emph{iff} every included $\underline{\mathbf{tx}}_1^i$ is valid, i.e.\ every paired $\underline{\mathbf{tx}}_0^i$ has been mined.
\end{description}

The benefit of this scheme is that it allows miners to amortize the cost of claiming transaction fees by not collecting them after each mined $\mathbf{tx}_0$, rather they can claim those accrued fees whenever they wish to collect them by simply closing the unidirectional payment channels.
In the payment-channel-based scheme, two extra on-chain transactions (for channel establishment and closure) are required for an arbitrary number of metatransactions.
On the contrary, in the miner-based scheme, for $N$ number of $\mathbf{tx}_0$, the miner has to issue $N$ on-chain $\mathbf{tx}_1$ accordingly to gather transaction fees.

\subsubsection{Satisfying the Security Goals}
To satisfy the security goals of metatransactions specified in Section~\ref{sec:securitygoals}, we emphasize the following atomicity conditions of our payment-channel-based scheme:
% \begin{itemize}
%     \item $\mathcal{M}$ will include $\mathbf{tx}^i_0$ in a block \emph{iff} the transaction fee $\underline{\mathbf{tx}}^i_{1,\mathbf{txFee}_{\mathcal{C}^{*}}}$ is secured once $\mathcal{M}$ mines $\mathbf{tx}^i_0$.
%     \item An off-chain transaction $\underline{\mathbf{tx}}_1^i$ in $C_{\mathcal{S}\rightarrow\mathcal{M}}$ is valid \emph{iff} $\mathbf{tx}^i_0$ has been mined by $\mathcal{M}$.
%     \item $\mathbf{tx}_{\sum_i}$ can be finanlized on-chain \emph{iff} every contained off-chain transaction $\underline{\mathbf{tx}}_1^i$ is valid, i.e. every on-chain transaction $\mathbf{tx}^i_0$ has been mined by $\mathcal{M}$
% \end{itemize}
\begin{condition}\label{condition:miningcondition}
$\mathcal{M}$ will include $\mathbf{tx}^i_0$ in a block \emph{iff} the transaction fee $\underline{\mathbf{tx}}^i_{1,\mathbf{txFee}_{\mathcal{C}^{*}}}$ is secured once $\mathcal{M}$ mines $\mathbf{tx}^i_0$.
\end{condition}

\begin{condition}\label{condition:offchaintxcondition}
An off-chain transaction $\underline{\mathbf{tx}}_1^i$ in $C_{\mathcal{S}\rightarrow\mathcal{M}}$ is valid \emph{iff} $\mathbf{tx}^i_0$ has been mined by $\mathcal{M}$.
\end{condition}

\begin{condition}\label{condition:aggregationtxcondition}
$\mathbf{tx}_{\sum_i}$ can be finanlized on-chain \emph{iff} every contained off-chain transaction $\underline{\mathbf{tx}}_1^i$ is valid, i.e. every on-chain transaction $\mathbf{tx}^i_0$ has been mined by $\mathcal{M}$.
\end{condition}

Thus $\mathcal{M}$ may be asked to proof on-chain that $\mathbf{tx}^i_0$ has been mined by $\mathcal{M}$ to make $\mathbf{tx}_{\sum_N}$ legitimate.
In most blockchains, transactions in a block are constructed in a Merkle tree (or a similar structure) with the Merkle root included in the block header.
As the miner of each block is verifiable, $\mathcal{M}$ can perform the proof by presenting Merkle tree inclusion proofs\footnote{By verifying Merkle tree or Merkle--Patricia trie inclusion proofs, depending what cryptographic accumulator scheme is implemented in the blockchain. See \url{https://github.com/lorenzb/proveth}}
and indicating which block contains this Merkle root.
However, in light of the expensive computation cost of on-chain Merkle tree proofs, we also allow $\mathcal{M}$ to provide the acknowledgement signed by $\mathcal{S}$ as the proof if $\mathcal{S}$ is cooperative.

In contrast to other payment channel designs (e.g.\ Lightning channels~\cite{poon2016bitcoin}, Duplex Micropayment Channels~\cite{decker2015fast}), Spilman unidirectional channels allow the sender to stay offline without the risk of losing funds~\cite{gudgeon2019sok}. Because of the unidirection, a rational miner only publishes the last aggregation transaction which pays the highest amount.
A miner must remain online to detect a timeout expiration for fraudulent channel closure.
In practice, we can expect a miner to remain online for the mining process.

Before such scheme is practical, we must solve the problem that the sender is not aware of which miner will mine $\mathbf{tx}_0$, which means the sender does not know a priori which miner to pay. Hence, the sender may uphold concurrently multiple open channels with different miners. Note, that when a sender issues off-chain payments to multiple miners, only one of these payments will be valid and redeemable by a miner. This is guaranteed by the transaction inclusion proof, the miner needs to provide when they redeem their earned metatransaction fees.

A miner adopting this scheme must monitor signed off-chain channel updates. Miners include transactions from the mempool \emph{iff} they received a newly updated channel balance referencing the transaction.

\section{Evaluation} \label{sec:comparingproposals}
In this section, we present our evaluation, compare the introduced metatransaction proposals qualitatively and assess quantitatively their induced overhead.

\subsection{Current Transaction Usage}
We empirically study the distribution of how transactions make use of the native currency in Ethereum. In Figure~\ref{fig:txratios}, we plot the gathered statistics taken from the Ethereum blockchain, from the genesis block up to block $\num[group-separator={,}]{8325000}$. We differentiate between $4$ classes of transactions: 

\begin{figure}%[11]{r}{0.65\textwidth}
\centering
\includegraphics[width=\linewidth,trim={6.3cm 2.4cm 2.5cm 0.5cm},clip]{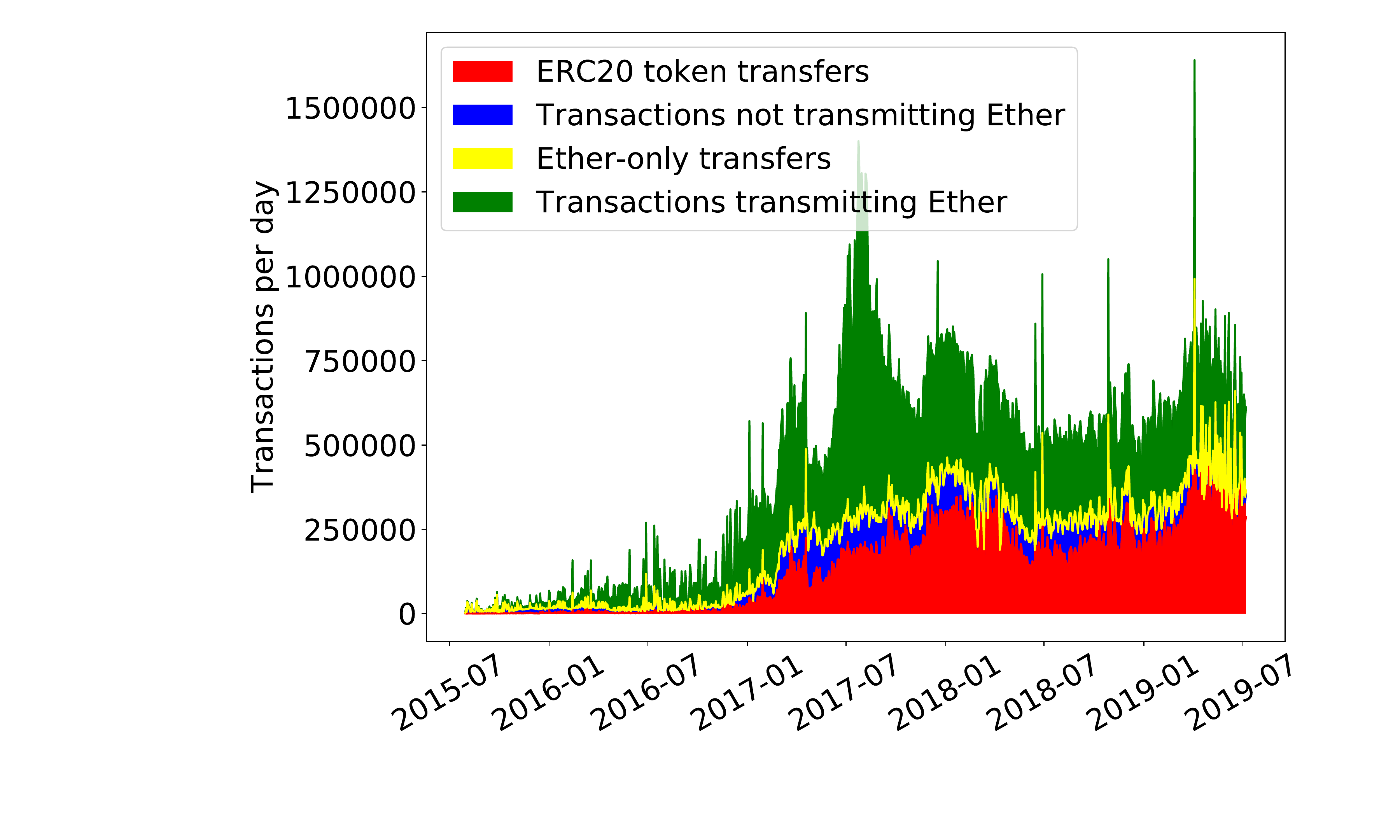}
\caption{We observe a gradual increase in the number of Ethereum transactions that involve the native currency only for paying transaction fees.}
\label{fig:txratios}
\end{figure}

\begin{description}
\item[Transactions transmitting Ether:] transactions that transfer Ether between accounts and may additionally perform other smart contract calls.
\item[Ether-only transfers:] a subset of the previous transaction class, in which transactions only transfer Ether and do not perform additional smart contract calls.
\item[Transactions not transmitting Ether:] transactions that do not transfer Ether.
\item[ERC20 token transfers:] a subset of the previous transaction class, in which ERC20 tokens are transferred.
\end{description}{}

We observe that on the 10th of August 2019, $46.9\%$ ($\num[group-separator={,}]{286736}$/$\num[group-separator={,}]{611467}$) of transactions performed were ERC20 transactions. The majority of transactions, $57.9\%$ ($\num[group-separator={,}]{354273}$/$\num[group-separator={,}]{611467}$), do not transfer Ether; they only involve Ether for paying transaction fees. Based on the findings in Figure~\ref{fig:txratios}, we conclude that in Ethereum, most transactions use the native currency exclusively to pay transaction fees. We empirically evaluate the number of transaction in Ethereum that do not pay any transaction fees. We do this to quantify the current adoption of metatransaction schemes.

\begin{figure}%[15]{r}{0.65\textwidth}
\includegraphics[width=\linewidth,trim={0.8cm 0cm 0cm 2.1cm},clip]{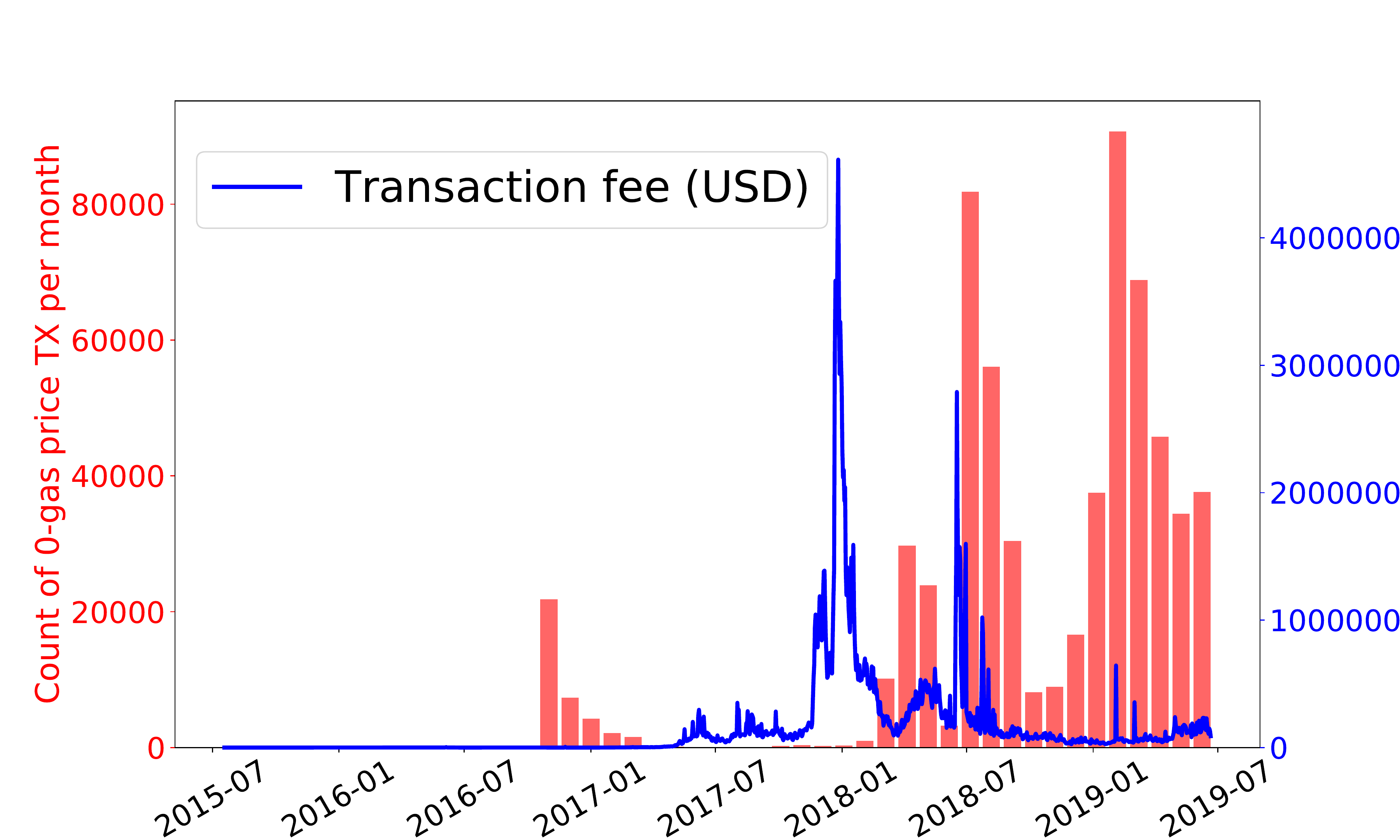}
\caption{The number of 0-gasprice transactions per month on the Ethereum blockchain correlates well with transaction fees.}
\label{fig:0gasprice}
\end{figure}

We found that the number of $0$-gasprice transactions correlates well with transaction fees, i.e., the Spearman correlation~\cite{lehman2005jmp} amounts to $\rho=0.5259$. This correlation indicates that whenever transaction fees are relatively high, miners seem to include their own transactions in their mined blocks to avoid paying high transaction fees. We note that $0$-gasprice transactions are unlikely to originate from regular users because $0$-gasprice transactions are not forwarded on the network layer.

\begin{algorithm}
\SetAlgoLined
   $\mathtt{broadcast}((s_0,r_{channel},0,collateral_{channel},\emptyset))$\;
   $senderBalance=\mathit{collateral}_{\mathit{channel}},counter=0$\;
   \While{$senderBalance\geq\mathbf{txFee}_{\mathcal{C}^{*}}\mathbf{and}\mathit{now}\leq \mathit{timeout}$}{
        $\mathtt{sendOffChain}((s_{channel},r_{miner},0,(counter+1)\cdot\mathbf{txFee}_{\mathcal{C}^{*}},\mathbf{tx^{counter}_{0}}))$\;
        $\mathbf{tx^{counter}_{0}}=(s_0,r_{0},0,0,\emptyset)$\;
        $\mathtt{broadcast}(\mathbf{tx^{counter}_{0}})$\;
        \If{$\mathit{mined}(\mathbf{tx^{counter}_{0}})$}{
        $senderBalance=senderBalance-\mathbf{txFee}_{\mathcal{C}^{*}}$\;
        $counter++$\;}
        
   }
   $\mathtt{broadcast}((s_{channel},r_{miner},0,collateral_{channel}-senderBalance,\emptyset))$\;
 \caption{A payment-channel-based metatransaction construction.  A sender creates a unidirectional payment channel with a miner locking up $collateral_{channel}$. Sender issues off-chain balance updates paying transaction fees in $\mathcal{C}^{*}$. Miner includes the referenced transaction in a subsequent block. They can keep open the channel until collateral is depleted or until a pre-determined timeout expires. For ease of exposition for each transaction we account a constant transaction fee $\mathbf{txFee}_{\mathcal{C}^{*}}$.}
 \label{alg:channelmetatransaction}
\end{algorithm}

\begin{figure*}
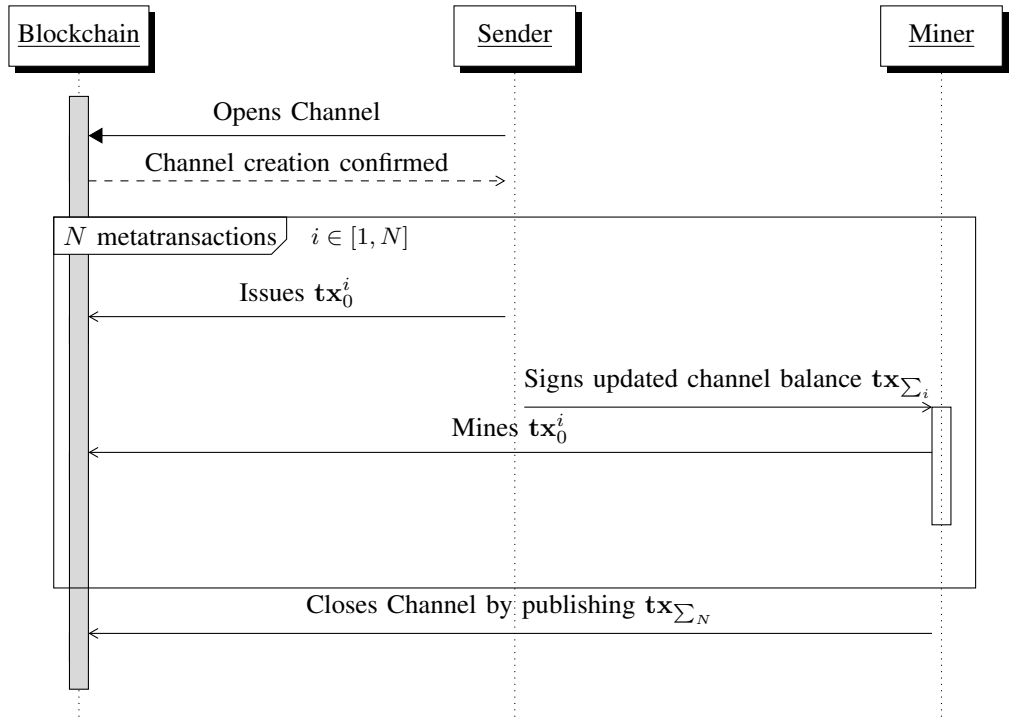

  \centering
  \begin{sequencediagram}
    \newthread{B}{Blockchain}{}
    \newinst[4]{A}{Sender}{}
    \newinst[4]{C}{Miner}{}
    \begin{call}{A}{Opens Channel}{B}{Channel creation confirmed}
    \end{call}
    \begin{sdblock}{$N$ metatransactions}{$i\in[1, N]$}
    \begin{messcall}{A}{Issues $\mathbf{tx}^i_0$}{B}{}
    \end{messcall}
    \begin{messcall}{A}{Signs updated channel balance $\mathbf{tx}_{\sum_i}$}{C}{
    \begin{messcall}{C}{Mines $\mathbf{tx}^i_0$}{B}
    \end{messcall}
    }
    \end{messcall}
    % \begin{messcall}{A}{Issues $\mathbf{tx}^{N}_0$}{B}{}
    % \end{messcall}
    % \begin{messcall}{A}{Signs updated channel balance $\mathbf{tx}_1$}{C}{
    % \begin{messcall}{C}{Mines $\mathbf{tx}^{N}_0$}{B}
    % \end{messcall}
    % }
    % \end{messcall}
    \end{sdblock}
    \begin{messcall}{C}{Closes Channel by publishing $\mathbf{tx}_{\sum_N}$}{B}{}
    \end{messcall}

  \end{sequencediagram}
  \caption{Sequence diagram of the payment-channel-based metatransaction construction. Sender opens a unidirectional payment channel with miner. Subsequently, a sender can issue multiple on-chain zero-fee transactions, while paying the corresponding transaction fee off-chain to a miner. Once the collateral of the channel is depleted or a miner chooses to collect fees, the miner closes the channel by signing and broadcasting the latest channel balance $\mathbf{tx}_1$.} \label{fig:sequencediagram}
\end{figure*}
%}

\subsection{Miner-based Metatransaction}
Account-based cryptocurrencies typically apply nonces to deter replaying transactions~\cite{wood2014ethereum}. Each account's nonce is incremented by one after every issued transaction. Transactions with invalid account nonces cannot be mined. Hence, in the context of metatransactions, account nonces can be used to facilitate the atomicity of miner-based metatransactions. Specifically, senders first issue $\mathbf{tx}_0$ and subsequently $\mathbf{tx}_1$ with an incremented nonce. Although $\mathbf{tx}_1$ pays the metatransaction fees to miner, solely $\mathbf{tx}_1$ cannot be mined. The $\mathbf{tx}_1$ transaction is invalid without the inclusion of  $\mathbf{tx}_0$ due to the applied nonce mechanism. The metatransaction fee in $\mathbf{tx}_1$ can be transferred to an anyone-can-spend address, which is later redeemed by the miner of the block.  

However, for example in Ethereum, one can access the miner's address of the current block due to the scripting language of the platform. Therefore there is no need to establish an anyone-can-spend address, i.e. senders can directly transfer metatransaction fees to miners in $\mathbf{tx}_1$. Therefore, in case of Ethereum, each issued miner-based metatransaction induces the overhead of an additional internal transaction. The gas overhead of $\mathbf{tx}_0$, the additional transaction, in case of Ethereum, amounts to a $\num[group-separator={,}]{15188}$ gas overhead, cf. Table~\ref{tab:gascosts}. 
%An implementation of the scheme for Ethereum, can be found in Appendix~\ref{sec:codes}.

We remark that unspent transaction output (UTXO)-based cryptocurrencies, like Bitcoin, can adopt the miner-based metatransaction scheme with minimal overhead. Such metatransaction only contains one additional UTXO, which essentially implements $\mathbf{tx}_1$. Namely, $\mathbf{tx}_1$ sends the metatransaction fee to an anyone-can-spend address, which can be redeemed by the miner of the block.

\subsection{Payment-channel-based Metatransaction}
Payment-channel-based metatransactions could potentially incur minimal overhead for the blockchain. Practically speaking, thousands of metatransactions can be sent only by issuing two on-chain transactions. Nevertheless, the on-chain footprint of this construction seems minimal as two on-chain transactions are necessary to establish and close the payment channel with the miner, cf. Algorithm~\ref{alg:channelmetatransaction}.

\begin{figure}%[15]{r}{0.65\textwidth}
\includegraphics[width=\linewidth,trim={0cm 0cm 0cm 0cm},clip]{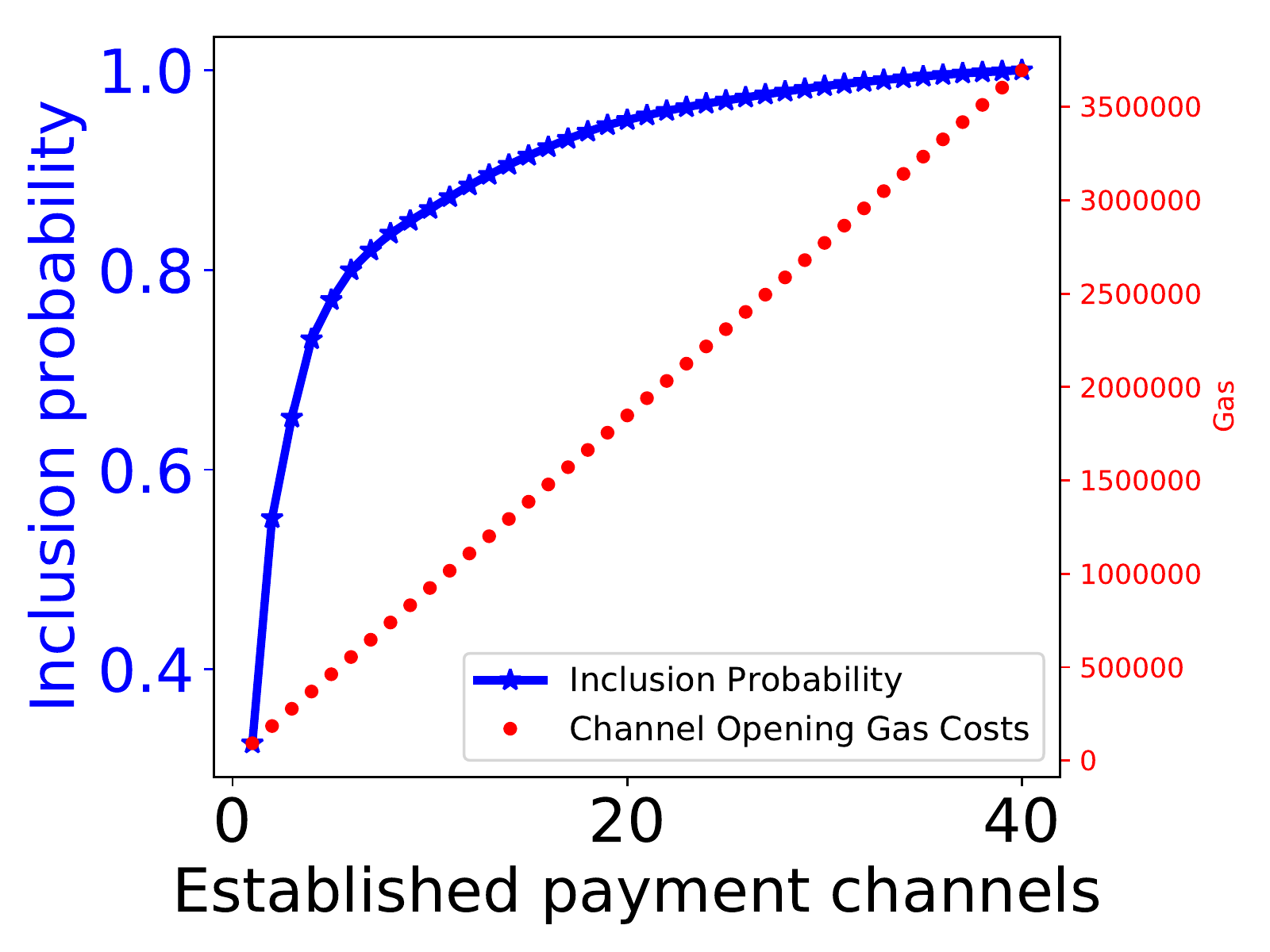}
\caption{Metatransaction inclusion probability as the function of established payment channels in the payment-channel-based metatransaction scheme.}
\label{fig:practicability}
\end{figure}

To show the practicability of this scheme we conduct several measurements. First, we note that as of 2020 March 22, in Ethereum, there are dozens of active mining pools (cca. 70)~\footnote{Source: \url{https://etherscan.io/stat/miner?blocktype=blocks}}. We observe that the hashrate distribution of mining pools follow an exponential distribution with $\lambda=2.4045$ ($\chi^2$ = 6.0053, p = 0.7393). This already suggests that a handful of established payment channels with miners might suffice to provide sufficiently fast transaction inclusion, cf. Figure~\ref{fig:practicability}. Indeed, five mining pools control as large as $75\%$ of the network's hashrate. Hence, as few as five established payment channels per user to these largest mining pools guarantees approximately $75\%$ transaction inclusion probability. The incurred gas cost of opening payment channels is linear in the number of established channels. Therefore, more opened payment channels yield only marginal gains in transaction inclusion probability as the hashrate of miners follow an exponential distribution.

\begin{figure}%[15]{r}{0.65\textwidth}
\includegraphics[width=\linewidth,trim={0cm 0cm 0cm 0cm},clip]{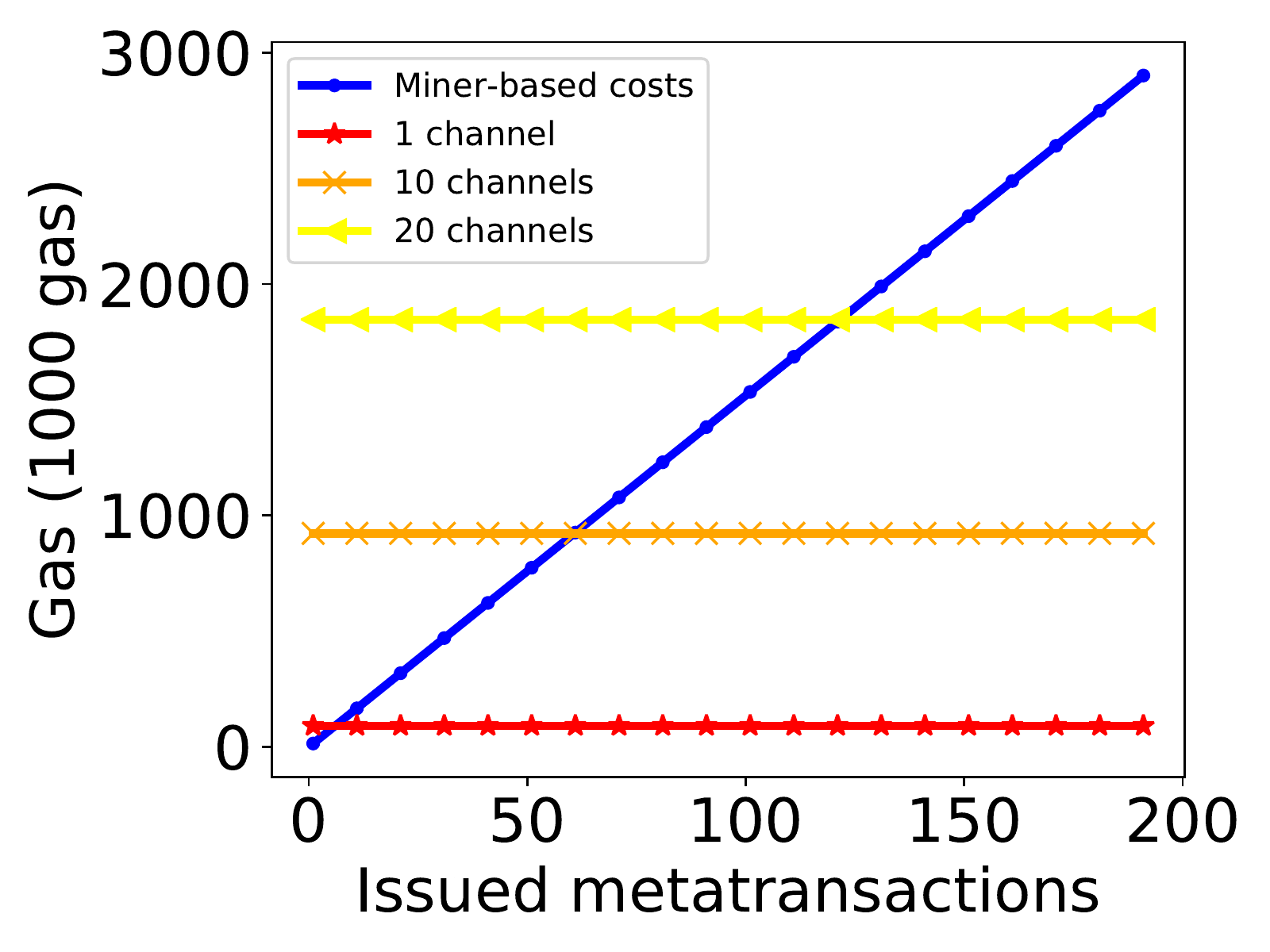}
\caption{Break-even points for the payment-channel-based metatransaction construction in comparison with the miner-based construction.}
\label{fig:breakeven}
\end{figure}

Repeated usage of metatransactions would render the payment-channel based construction more practical than the miner-based construction, cf. Figure~\ref{fig:breakeven}. In the miner-based construction, each issued metatransaction produces a constant, $15,188$ gas overhead (cf. Table~\ref{tab:gascosts}), while the payment-channel based construction incurs no additional gas overhead. Hence, repeated metatransaction usage amortizes the cost of opening channels with miners. In case of Ethereum, the cost of opening a payment channel is $92,392$ gas. Therefore, if a user issues at least $7$, $61$, $122$ metatransactions, then they can amortize the cost of opening $1$, $10$ and $20$ payment channels with miners, respectively. Nonetheless, whenever a user wants to send $\mathit{cnt}$ transactions and assuming constant metatransaction fees, $\mathit{txFee}$, then the initial deposit amounts to $\mathit{cnt}\cdot\mathit{txFee}\cdot N$, where $N$ is the number of opened payment channels to mining pools. Accordingly, requiring initial deposits from users generates opportunity cost, hence constitutes a drawback of the payment-channel based construction. 
However, we remark that if we assume a dense payment channel network between users and miners, then users might reduce their initial deposit costs as they could route and rebalance their discharged payment channels using others' non-discharged channels~\cite{khalil2017revive}. Consequently, an in-depth practicability and utility analysis of the payment-channel based scheme is a complex problem, which we leave for future work.

The closure of the payment channel may be computationally lightweight in the graceful case, i.e. when there is no dispute between miner and sender. Although in the worst case, miners need to include transaction inclusion proofs, whose added overhead may vary quite widely from $\num[group-separator={,}]{850000}$ to millions of gas, in case of Ethereum, cf. Table~\ref{tab:gascosts}. 
%An implementation of the scheme is enclosed in Appendix~\ref{sec:codes}.

\subsection{Fee Delegation vs.\ Metatransactions}
We compare the presented metatransaction and fee delegation scheme considering the following properties:

\begin{description}
\item [Non-custodial] Do senders remain the custodians of their assets?
\item [Liveness] Do sender rely on an online third party besides the underlying blockchain? 
%(e.g., relayer network or individual relayers)
\item [Multiple Delegators] Does the solution support multiple delegators/relayers?
\item [Integration] Can a smart contract application adopt the metatransaction protocol without code changes?
%\item [Cost] Does the metatransaction mechanism increase or decrease the transaction fees compared to a payment in the native cryptocurrency?
%\item [Rate Limiting] {\color{red}don't understand:}Are there constraints on how many fees can be delegated per user?
%\item [Versatility] {\color{red}don't understand:}Can this implementation work for many use cases or just one?
%\item [Payee Flexibility] {\color{red}don't understand:}Can the payee be anyone, or is it just the creator of the contract, etc?
\item [Censorship resistance] Can transactions be censored?
\end{description}

\begin{table}
  \centering
    \label{tab:table1}
    \begin{tabular}{lccc} \toprule
       & \shortstack{\textbf{Miner}\\Section~\ref{sec:minerscheme}} & \shortstack{\textbf{Relayer}\\Section~\ref{sec:relayerschemes}} & \shortstack{\textbf{Channel}\\Section~\ref{sec:channelscheme}}\\
        \hline
      Non-custodial & yes  & yes& yes\\
      Liveness & no & yes& no\\
      Multiple delegators & yes & no & yes\\
      Integration & no & no& no\\
      %Cost & low & medium& low&high&high\\
      %Rate Limiting & yes & yes& yes&yes&yes\\
      %Versatility & yes& yes& yes&yes&yes\\
      %Payee flexibility & flexible & fixed/(flexible) & flexible&flexible&flexible\\
      Censorship resistance & yes & no & yes\\
      \bottomrule
    \end{tabular}
    \caption{Comparing qualitatively two metatransaction and one fee delegation scheme.}
\end{table}

To quantitatively evaluate the practicality of metatransactions we measure the incurred overhead of the discussed proposals in terms of gas costs and the number of additional transactions in Table~\ref{tab:gascosts}. Our experimental setup is structured as follows. On a local Ethereum private network (geth v1.8.22.), we deployed an ERC20 token contract with extended functionalities to enable metatransactions. For the miner and channel-based metatransactions we created a separate function on the ERC20 token contract (Solidity version 0.5) that performs the reimbursement of a miner. We quantify the overhead of issuing a metatransaction with the execution of these functions. Therefore we measured the gas costs of calling these functions\footnote{EIP-1108, \url{https://github.com/ethereum/EIPs/blob/master/EIPS/eip-1108.md} is expected to render the private (meta)transaction proposals~\cite{bunz2019zether,williamson2018aztec} significantly cheaper.}. We observe that the miner-based scheme is the most lightweight followed by the relayer-based.

\begin{table}
  \begin{center}
    \label{tab:table1}
    \begin{tabular}{lccc}\toprule % <-- Alignments: 1st column left, 2nd middle and 3rd right, with vertical lines in between
       & \textbf{Miner} & \textbf{Relayer} & \textbf{Channel}\\
      \hline
      Tx per Block & $k$ internal tx & $k$ internal tx& $0^{*}$\\
      Gas per Tx & $\num[group-separator={,}]{15188}$& $\num[group-separator={,}]{47241}$& $0^{*}$\\
      \bottomrule
    \end{tabular}
    
  \end{center}
  \caption{Added overhead to regular Ethereum transactions in terms of incurred gas costs and additional issued transactions for the various metatransaction proposals. We denote the number of metatransactions included in a block with $k$. All metatransaction schemes require $k$ additional internal transactions, except the channel-based construction. Note that we omit the channel-opening ($ \num[group-separator={,}]{250000}$ gas) and closing ($\num[group-separator={,}]{850000}$ gas) costs, since they are amortized over the course of numerous issued metatransactions. Using the channel construction does not incur on-chain fees since they are paid off-chain.}\label{tab:gascosts}
\end{table}

\begin{figure}%[18]{l}{0pt}
\centering
    \includegraphics[width=0.7\linewidth,trim={0cm 0cm 0cm 0cm},clip]{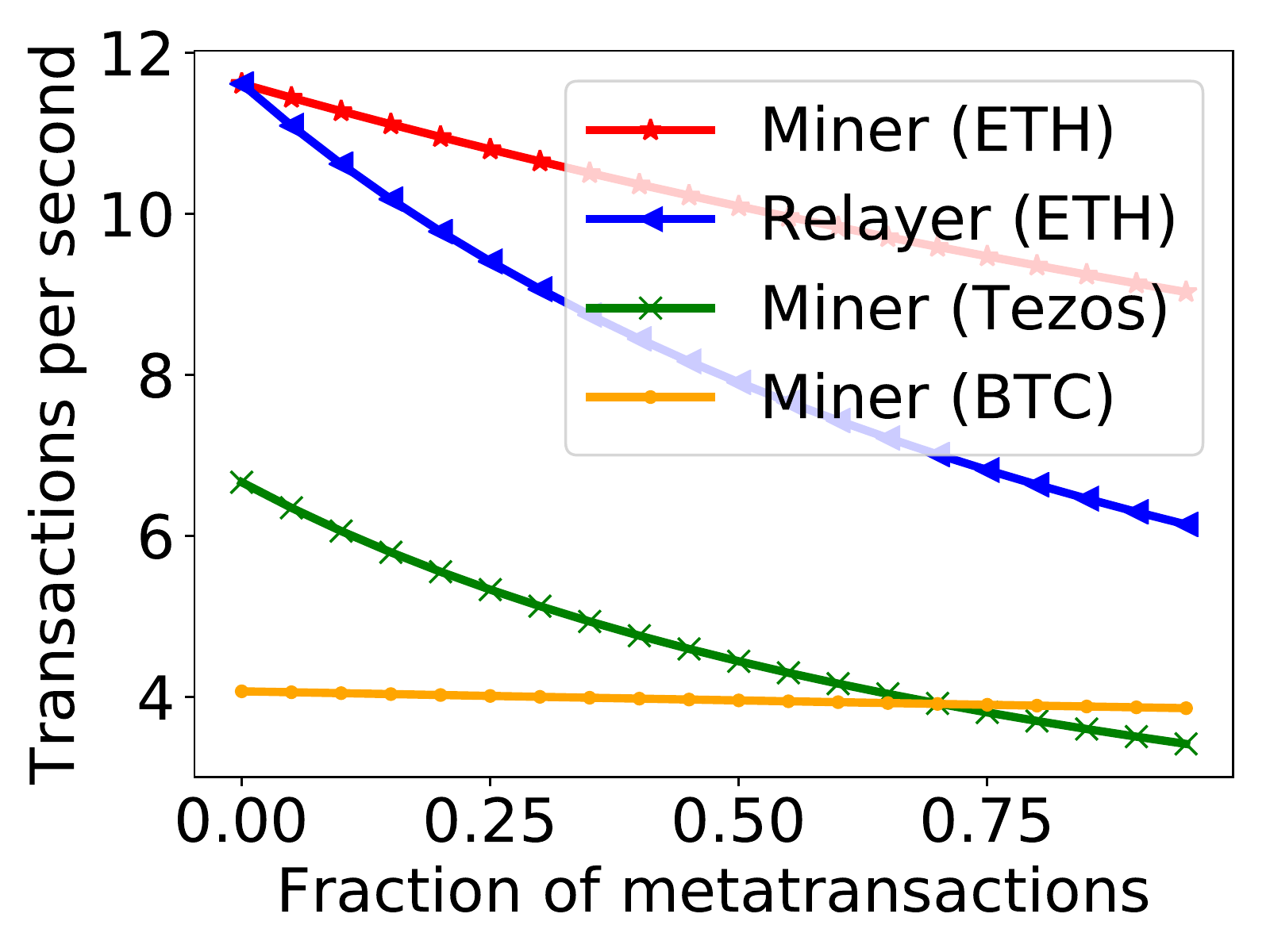}
    \caption{Throughput analysis of various metatransaction and fee delegation proposals. Miner-based construction yields the highest throughput, followed by the relayer-based metatransaction scheme. For each blockchain and proposal we calculated the system's throughput as a function of the fraction of metatransactions among all issued transactions on the blockchain.}\label{fig:throughput}
\end{figure}{}
We measure the caused impact on Bitcoin's, Ethereum's and Tezos' throughput to assess the economic viability of metatransaction proposals. When the fraction of metatransactions increases, throughput monotonically decreases.  

%https://explorer.bitcoin.com/bch/charts BCH tps
In case of Bitcoin, one can implement the miner-based protocol using a recent proposal\footnote{See: \url{https://simpleledger.cash}}. Any transaction can implement token transfers in Bitcoin by adding a single 0-value coin (unspent transaction output) to the transaction data. The average size of an unspent transaction output (UTXO) amounts to $93.45$ bytes~\cite{delgado2018analysis}. On October 21, 2019, $\num[group-separator={,}]{351791}$ transactions were appended to the blockchain with an average 1637 byte size per transaction. We observe that if all transactions in Bitcoin would apply the miner-based metatransaction scheme, then the throughput of the blockchain would decrease from $4.07$ transaction per second to $3.86$ transaction per second, cf. Figure~\ref{fig:throughput}.

For Ethereum and Tezos, we calculated with a $\num[group-separator={,}]{8003131}$ and $\num[group-separator={,}]{4000000}$ block gas limit\footnote{Source: \url{https://etherscan.io/chart/gaslimit}} and with a $11.62$ and $6.67$ transaction per second for Ethereum and Tezos respectively, as observed on July 16, 2019. Added gas overheads per metatransaction were taken from Table~\ref{tab:gascosts}. Ethereum's throughput decreases by $48.13\%$ and $22.97\%$ if all transactions in a block adapt the relayer fee delegation and miner metatransaction schemes respectively, cf. Figure~\ref{fig:throughput}. In case of Ethereum, the miner-based scheme in a non-backward compatible way only requires a single token transfer to the miner's address. On the other hand, Tezos currently only supports this metatransaction protocol ($\num[group-separator={,}]{10000}$ gas per transaction) with the difference that there one needs to apply an anyone-can-spend address as the miner's address of a block is not available in the scripting language. The relayer-based fee delegation scheme is slightly more computationally involved. %For the cross-chain scheme, we calculated with the Summa-proofs cost~\cite{prestwich2018summa}, and for the private metatransaction protocol, we used the gas cost of a 1-to-2 join-split transaction from the AZTEC protocol~\cite{williamson2018aztec}, cf. Table~\ref{tab:gascosts}.
Overall we found that miner-based metatransactions incur the least overhead.

\par\smallskip
\noindent\textbf{Deploying metatransactions} \label{sec:deployingmetatx}
In most blockchain P2P networks, $0$-fee transactions are considered as a DoS attack and are not forwarded among peers. Therefore, we believe that the majority of 0-gasprice transactions we have witnessed in Figure~\ref{fig:0gasprice} are originating from miners. To facilitate metatransactions peers could allow those $0$-gasprice transactions which reimburse miners in non-native currencies. This added verification would yield insignificant overhead both for miners and other full nodes. Furthermore, obstacles to deploying metatransactions are that miners might not accept tokens used for metatransactions as legal tender. Negotiating a suitable token as a payment method can be cumbersome and challenging. 

\section{Economic and security implications of metatransactions} \label{sec:securityofmetatx}

Here we discuss the economic and security impact of metatransactions. We describe one possible avenue to perform a malicious takeover of a blockchain's native currency.

\subsection{Discussing the economic impact of metatransactions} 
In traditional macroeconomics, participants of an economy typically look for substitutes of their own currency when it is sharply devalued or inflated. Most commonly these substitutes are more stable currencies of other countries~\cite{piontkovsky2003dollarization}. On the other hand, in cryptoeconomics, most of the cryptocurrencies exhibit immense volatility~\cite{dyhrberg2016bitcoin}, while there are also available deployed stablecoins~\cite{team2017dai}. Therefore cryptocurrencies seem like an ideal setting for traditional currency substitution, in our jargon metatransactions. Hence by applying the theory of currency substitution to the field of cryptoeconomics, one can expect cryptocurrency users to substitute their volatile, unstable native cryptocurrencies (e.g. ether) with stablecoins (e.g. DAI) issued on top of their used cryptocurrency platform.

Metatransactions would decrease the demand for a native currency while likely increasing demand for other currency $\mathcal{C}^{*}$. For instance, if all transactions are metatransactions and paid in the stablecoin DAI~\cite{team2017dai}, then metatransaction fees paid in DAI would have increased DAI daily volume\footnote{Source: \url{https://coinmarketcap.com/currencies/dai/historical-data/}} by a minimum of $1.02\%$ up to $3.26\%$ as observed on August 13, 2019, and  February 19, 2019, respectively. Additionally, senders are more likely to use other currencies if their native currency's inflation rate increases~\cite{drenik2017pricing}.

\subsection{Quantifying the security impact of metatransactions}

To measure the security of blockchains equipped with metatransactions, we apply the quantitative framework introduced in~\cite{gervais2016security}. Namely, we adopt the value of double-spending, $v_d$, as a generic metric for the security of blockchains. More precisely, $v_d$ denotes the minimal double-spending value, which is strictly larger than the honest mining reward: 
\begin{equation}{}
v_d=min\{v_d|\exists \pi \in A: R(\pi,P,v_d)> R(honest\; mining, P) \},\label{eq:vddef}
\end{equation}
where $R$ is the reward matrix, $A$ is the action space, $P$ is the stochastic transition matrix and $\pi$ is a policy. To find the optimal policy $\pi$ we use the Markov-Decision Processes (MDP). 

\begin{figure}%[21]{l}{0pt}
\centering
    \includegraphics[width=0.8\linewidth,trim={3cm 2cm 3cm 1cm},clip]{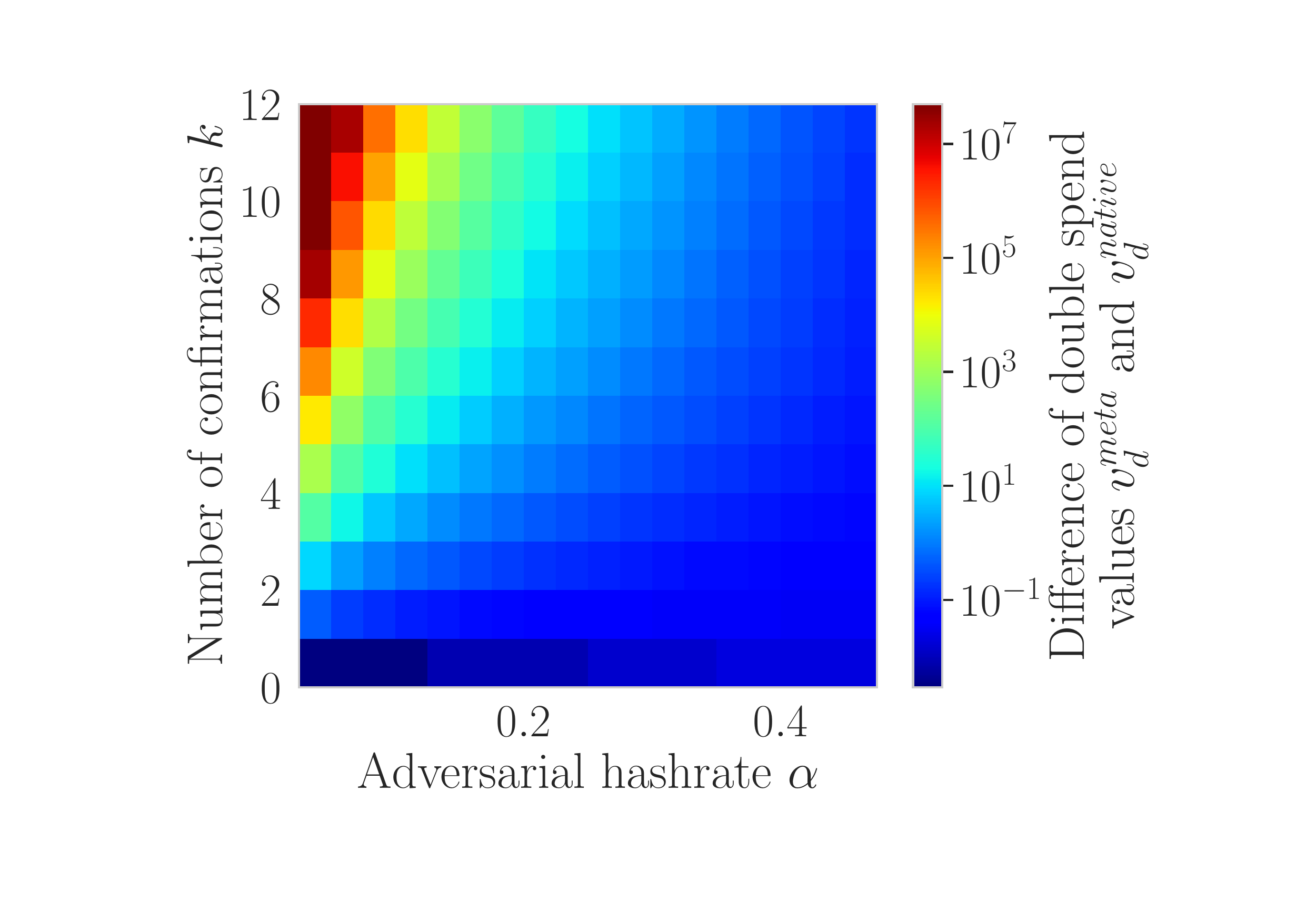}
    \caption{Difference of the security values $v_{d}$ of Ethereum, when it is equipped with metatransactions, $v_{d}^{\mathit{meta}}$, and when it is not, $v_{d}^{\mathit{native}}$. Security is expressed as  the minimal double-spending value $v_{d}$, which is strictly larger than the honest mining reward. Longer confirmations and less potent malicious miners increase security. }
    %\caption{Security of Ethereum equipped with metatransactions. Security is expressed as  the minimal double-spending value denoted as $v_d$, which is strictly larger than the honest mining reward. Longer confirmations and less potent malicious miners increase security. }
    \label{fig:securityMetaTx}
\end{figure}

We assume a rational adversary that is interested in maximizing his benefits, measured in financial gains, in the network. We extend the MDP of~\cite{gervais2016security}, to model (meta)-transaction fees as well. We model (meta)-transaction fees as added constants on top of block rewards. For instance, as of writing, in case of Ethereum approximately $5\%$ of the full block reward comes from transaction fees. Furthermore, we assume that the metatransaction token is as liquid as the native currency. We leave it as a fascinating future work to model dynamically changing (meta)transaction fees as well. We observe that there is no major difference between the security of blockchains either they apply metatransactions or not in the constant transaction fee model, see Figure~\ref{fig:securityMetaTx}. Therefore the introduction of metatransactions does not negatively impact the security of a PoW blockchain. Another interesting avenue for future work would be to extend the MDP to allow modelling of bribing attacks. Double-spender adversaries might incentivise miners to mine on top of their branch by offering large (meta)transaction fees to them~\cite{liao2017incentivizing}.

\subsection{Blockchain currency takeover} \label{sec:currencytakeover}

One possible application of metatransactions is the potential replacement of the native-currency $\mathcal{C}$ for paying transaction fees. 
An attacker motivated by making a certain currency $\mathcal{C}^{*}$ dominant can displace the native currency  of any blockchain by subsidizing transactions offering transaction fees in $\mathcal{C}^{*}$. For each transaction $\mathbf{tx}$ offering metatransaction fee ${\mathbf{txFee}}_{\mathcal{C}^{*}}$, the attacker reimburses transaction sender with an equal amount in $\mathcal{C}^{*}$. From a sender perspective, the attack essentially results in feeless transactions. Such a currency takeover can be launched trustlessly by a token contract and it incentivises sender to no longer use $\mathcal{C}$ as a transaction fee currency.
\begin{table}
  \begin{center}
    \begin{tabular}{lccc} % <-- Alignments: 1st column left, 2nd middle and 3rd right, with vertical lines in between
    \toprule
       & \textbf{$1$ hour} & \textbf{$1$ day} & \textbf{$1$ week}\\
      \hline
      $51\%$ attack & \textdollar $\num[group-separator={,}]{360114}$ &\textdollar $\num[group-separator={,}]{8642736}$  & \textdollar $\num[group-separator={,}]{60499152}$\\
      Currency takeover & 
      \textdollar $\num[group-separator={,}]{8290}$ &\textdollar $\num[group-separator={,}]{198963}$  & \textdollar $\num[group-separator={,}]{1392740}$\\
      \bottomrule
    \end{tabular}
  \end{center}
  \caption{The financial costs of launching a $51\%$ attack and a currency takeover on the Ethereum blockchain. }\label{tab:currencytakeovercosts}
\end{table}

A blockchain currency takeover may be executed as follows:
\begin{enumerate}
	\item The attacker deploys the $\mathit{CurrencyTakeover}$ smart contract and advertises its address. The contract issues a non-native currency $\mathcal{C}^{*}$, which may be used to pay transaction fees. 
	\item Sender can issue metatransactions $\mathbf{tx}_{meta}=(\mathbf{tx}_0,\mathbf{tx}_1),$ where $\mathbf{tx}_1$ calls the $\mathit{CurrencyTakeover}$ contract and transfers transaction fees $\mathbf{tx}_{1,\mathbf{txFee}_{\mathcal{C}^{*}}}$ in $\mathcal{C}^{*}$ to miners.
	\item Whenever sender issue metatransactions and $\mathbf{tx}_1$ is executed, senders are automatically and trustlessly reimbursed. Particularly, sender are refunded with the same amount, they have just sent to miners in transaction fees, i.e. $\mathbf{tx}_{1,\mathbf{txFee}_{\mathcal{C}^{*}}}$ in $\mathcal{C}^{*}$. Thereby, due to the attacker, sender effectively spend no transaction fees if they use the currency $\mathcal{C}^{*}$ to pay the transaction fees.
\end{enumerate}

To put the financial costs of a currency takeover into perspective we compare it with a rental $51\%$ attack, where the attacker rents mining hardware, cf. Table~\ref{tab:currencytakeovercosts}. Renting an NVIDIA K80 GPU Ethereum miner on Amazon’s Elastic Compute Cloud (EC2) platform costs  US$\$0.20$ per hour and it is able to perform $24$MH/s~\cite{bonneau2018hostile}. As of 16 September 2019 the Ethereum network hashrate was approximately $2.2$TH/s~\footnote{Source: \url{https://etherscan.io/chart/hashrate}}. On the same day, we observed a $US\$197.11$ Ether price and that $\num[group-separator={,}]{1009.4}$ Ether was paid for transaction fees.

We remark that similar to the rental $51\%$ attack, also the currency takeover is temporary unless wallets and full nodes support metatransactions. The attack stops once the attacker does not hold funds to reimburse transaction sender. Furthermore, note, that currency takeover is a milder attack as it does not allow double-spending.

\subsubsection{Subsidy pool maintenance}
As mentioned above, a crucial aspect of the currency takeover is to maintain and sustain the subsidy pool. This is essential for the adversary in order to uphold their currency takeover. If the subsidy pool was depleted, the attacker could not subsidise users anymore to send metatransactions, hence users would not have additional financial incentives to issue metatransactions. Let $X$ denote the metatransaction fee of a transaction issued by a user and let $f(X)$ be the subsidy offered by the attacker after every issued metatransaction. In the following, we consider two cases.

\begin{itemize}[leftmargin=*]
    \item $f(X)=X$. Naively, an adversary might set $f(\cdot)$ to be the identity function. In this case, a counter-attacker, e.g. a miner or a wealthy entity, could rapidly deplete the subsidy pool just by issuing metatransactions with large fees. Note, that in this case, the counter-attack is essentially free. In summary, the problem with the identity function is that transaction senders are not incentivised to issue transactions with their true valuation of the transaction fee. However, this can be amended.
    \item $f(X)=\sqrt{X}$. To avoid warped incentives in the transaction fee valuation, one could apply another family of monotonically increasing functions. We propose the $\sqrt{\cdot}$ function since there exists a well-known incentive-compatibility theorem from the quadratic voting literature~\cite{lalley2018quadratic}. One might think of $f(X)$ as the number of votes and $X$ as the cost of acquiring $f(X)$ votes. It was shown by Lalley and Weyl~\cite{lalley2018quadratic} that the equilibrium is when users put their true valuation of $X$, as they only have a marginal increase in their acquired votes (in our case metatransaction fee subsidy). Put differently, this reward function achieves that users will be incentivised to apply transaction fees they would use under regular circumstances. Hence, they will not be incentivised to counter-attack and deplete the subsidy pool. 
\end{itemize}{}

\section{Related Work}\label{sec:related-work}
We build upon a rich body of literature understanding how transaction fees affect the security provisions of cryptocurrencies. Whale transactions are transactions with an anomalously large transaction fee to convince miners to fork the current chain~\cite{liao2017incentivizing}. Bonneau~\emph{et al.} introduce the notion of bribery attacks~\cite{bonneau2016buy}, which are conceptually related as in a bribery attack an attacker incentivises other miners to mine on a fork preferred by the attacker. In hostile blockchain takeovers, an attacker with an extrinsic motivation disrupts the consensus process~\cite{bonneau2018hostile}. Similar in spirit, we devise a blockchain currency takeover attack to replace the native currency of a blockchain as transaction fee currency.
Möser and Böhme were the first to empirically analyse Bitcoin's nascent transaction fee market~\cite{moser2015trends}. Kroll~\emph{et al.}~\cite{kroll2013economics}, Houy~\cite{houy2014economics}, and Kaskaloglu~\cite{kaskaloglu2014near} consider the economics of Bitcoin transaction fees in the presence of adversaries. Additionally, they discuss potential changes to transaction fees and their policies. Easley~\emph{et al.} assert that Bitcoin without transaction fees is not viable~\cite{easley2019mining}. Even though Bitcoin block rewards steadily decrease and the transaction fee market became more efficient and mature, Carlsten~\emph{et al.} show that no cryptocurrency remain stable in a block reward-free regime~\cite{carlsten2016instability}. We remark that all these works studied a single-currency setting. To the best of our knowledge, we are the first to study metatransactions and assess their security provisions. 

Furthermore, we also consider the economic literature about currency substitution and multi-currency economies as related work as they also investigate and model the impact of introducing multiple currencies as legal tender in a potentially open economy~\cite{akccay1997currency,drenik2017pricing}. More volatile exchange rates are predicted in multi-currency settings and sender are anticipated to adopt less inflated currencies. Nonetheless, these results might be applicable only partly as economic literature assumes the existence of developed economies behind currencies, while cryptocurrencies generally lack fully-fledged economies, most of them are still only considered as speculative assets.

\section{Future directions}\label{sec:extensions}
\par\smallskip
\noindent\textbf{Metacoinbase}
So far, we only considered metatransactions to pay for transaction fees. With the flexibility of smart contract based blockchains, however, one could design a new block reward atop an existing blockchain, which is paid out in a currency $\mathcal{C}^{*}$. Such \emph{metacoinbase} reward could be paid out to a miner under specific conditions (e.g., the reward is only paid if the miner did not accept $\mathcal{C}$ as a currency to pay transaction fees). Carlsten et al.~\cite{carlsten2016instability} show that cryptocurrencies do not remain stable in a block reward-free regime, further motivating the introduction of new block rewards.

\par\smallskip
\noindent\textbf{Cross-chain Metatransactions}\label{sec:xchainscheme}
A sender on blockchain $\mathcal{B}_0$ might want to pay transaction fees to miners in currency $\mathcal{C}_1$ of blockchain $\mathcal{B}_1$. This would require for a miner in $\mathcal{B}_0$ to be aware of $\mathcal{B}_1$, and prove to $\mathcal{B}_1$ that the sender's transaction in $\mathcal{B}_0$ was indeed included in the blockchain. Different techniques might allow performing such constructions~\cite{bunz2019flyclient,kiayias2017non,prestwich2018summa}, we, however, leave further details for future work.
%if and only if a miner can prove that on blockchain $\mathcal{B}_0$ they included a certain transaction. The sender can verify the NIPoPoW on blockchain $\mathcal{B}_1$ attesting to the claim that transaction is included in a block mined by the prover on blockchain $\mathcal{B}_0$. 
%Non-Interactive Proofs of P in~\cite{bunz2019flyclient}.
%One can apply Summa proofs~\cite{prestwich2018summa} to ensure constant-sized proofs. Summa proofs only sample the last $k$ blocks of the blockchain and not a poly-logarithmic sample of it. This efficiency gain allows a smart contract verifying the proof today, however, {\color{blue}at the cost of sacrificing security} \hl{(first half of this sentence was going well, but then it seemed to lose coherence)}: Summa proofs are shown to be not composable~\cite{zindros2019summa}. This might not be a problem if a sender ensures that the value of her transaction is smaller than the cost of forking the blockchain.

\par\smallskip
\noindent\textbf{Private Fee Auction}\label{sec:privacyenhancedmetatx}
Current blockchain transaction fee designs correspond to a public auction format. The transactions of a sender that pays the most fees are likely included first in a blockchain. Our payment-channel based metatransaction proposal (cf.\ Section~\ref{sec:channelscheme}) is to our knowledge the first scheme which allows hiding transaction fees from competing sender. The consequences of such private auction mechanism would be interesting to study in future work.

\bibliographystyle{splncs04}
\bibliography{metatransactions}

\end{document}